\documentclass[twocolumn,showpacs,preprintnumbers,amsmath,amssymb]{revtex4}

\usepackage{graphicx}
\usepackage{dcolumn}
\usepackage{bm}

\begin{document}

\title{On Optimality Condition of Complex
Systems: Computational Evidence\\ }

\author{Victor Korotkikh}
\email{v.korotkikh@cqu.edu.au}
\author{Galina Korotkikh}
\email{g.korotkikh@cqu.edu.au} \affiliation{Central Queensland
University, Mackay, Queensland 4740, Australia}

\author{Darryl Bond}
\email{dbond@nrggos.com.au} \affiliation{ Central Queensland
University, Gladstone, Queensland 4680, Australia}

\begin{abstract}

A general condition determining the optimal performance of a
complex system has not yet been found and the possibility of its
existence is unknown. To contribute in this direction, an
optimization algorithm as a complex system is presented. The
performance of the algorithm for any problem is controlled as a
convex function with a single optimum. To characterize the
performance optimums, certain quantities of the algorithm and the
problem are suggested and interpreted as their complexities. An
optimality condition of the algorithm is computationally found: if
the algorithm shows its best performance for a problem, then the
complexity of the algorithm is in a linear relationship with the
complexity of the problem. The optimality condition provides a new
perspective to the subject by recognizing that the relationship
between certain quantities of the complex system and the problem
may determine the optimal performance.

\end{abstract}


\maketitle

The efficient management of complex systems is increasingly
important. However, despite significant progress and interest in
complex systems \cite{Holland_1}, there is a limited understanding
of the problem. In particular, because the existence of principles
governing the non-equilibrium situation has not yet been
established \cite{Ball_1}, the possibility of a general condition
determining the optimal performance of a complex system is still
unknown. We aim to contribute in this direction and present
results of computational experiments in order to discuss the
possibility of an optimality condition of complex systems. The
results provide a new perspective to the subject by recognizing
that the relationship between certain quantities of the complex
system and the problem may determine the optimal performance.

In this Letter, we consider an optimization algorithm as a complex
system by using the benchmark traveling salesman problems
\cite{Reinelt_1}. For any problem tested, we can control the
performance of the algorithm as a convex function with a single
optimum. Consequently, we take this opportunity to investigate
whether the performance optimums may be characterized in terms of
an optimality condition of the algorithm. Namely, we describe the
algorithm by a trace of the variance-covariance matrix derived
from the dynamics of the algorithm as a complex system. We also
specify a problem by a trace of the distance matrix. Remarkably,
the computational analysis of the performance optimums reveals a
relationship between the quantities of the algorithm and the
problem by approximating it well enough with a linear function.

To recognize the potential importance of the result, we interpret
the quantities as the complexities of the algorithm and the
problem, and formulate the following optimality condition: {\it if
the algorithm shows its best performance for a problem, then the
complexity of the algorithm is in the linear relationship with the
complexity of the problem}. Further experimental facts suggest how
the optimality condition may be extended into a wider context.

Let us consider an optimization algorithm ${\cal A}$ as a complex
system of $N$ computational agents that minimize the average
distance in the traveling salesman problem. All agents start in
the same city and at each step an agent visits the next city by
using one of two strategies: a random strategy, i.e., visit the
next city at random, or the greedy strategy, i.e, visit the next
closest city. We define that all agents start with the random
strategy. The state of the agents visiting $n$ cities of the
problem can be described at step $j = 1,...,n-1$ by a binary
sequence $s_{j} = s_{1j}...s_{Nj}$, where $s_{ij} = +1$, if agent
$i = 1,...,N$ uses the random strategy and $s_{ij} = -1$, if the
agent uses the greedy strategy to visit the next city. The
dynamics of the agents is realized through their choice of
strategies and can be encoded by an $N \times (n-1)$ binary
strategy matrix $S = \{ s_{ij}, \ i = 1,...,N, \ j = 1,...,n-1 \}
$.

We introduce a variable parameter $v$ that controls the dynamics
of the agents in a specific manner. Let $D_{ij}$ be the distance
travelled by agent $i=1,...,N$ after $j=1,...,n-1$ steps and
$$
D_{j}^{-}=\min_{i=1,...,N}D_{ij}, \
D_{j}^{+}=\max_{i=1,...,N}D_{ij}.
$$
All distances travelled by the agents after $j = 1,...,n-1$ steps
belong to the interval $[D_{j}^{-},D_{j}^{+}]$. The parameter $v$
specifies a threshold point
$$
D_{j}(v)=D_{j}^{+}-v(D_{j}^{+}-D_{j}^{-}), \ \ \  0 \leq v \leq
1+\gamma, \ \ \ \gamma = 0.1
$$
dividing the interval into two parts, i.e., successful
$[D_{j}^{-}, D_{j}(v)]$ and unsuccessful $(D_{j}(v),D_{j}^{+}]$.
If the distance $D_{ij}$ travelled by agent $i = 1,...,N$ after
$j=1,...,n-1$ steps belongs to the successful interval, then we
regard the agent's last strategy as successful. If the distance
$D_{ij}$ belongs to the unsuccessful interval, then the agent's
last strategy is unsuccessful.

In order to choose the next strategy, each agent uses an optimal
rule \cite{Korotkikh_1} that relies on the Prouhet-Thue-Morse
(PTM) sequence and has the following description:

1. {\it If your last strategy is successful, continue with the
same strategy.

2. If your last strategy is unsuccessful, consult PTM generator
which strategy to use next}.

The PTM sequence $+1-1-1+1-1+1+1-1 \ ... \ $ gives a symbolic
description of chaos resulting from period-doubling
\cite{Feigenbaum_1} in complex systems \cite{Allouche_1}. It can
be also associated with formation processes of integer relations
progressing well through the levels of a hierarchical structure
\cite{Korotkikh_1}. The formation processes produce a measure of
complexity \cite{Korotkikh_1} that we use as a guide in defining
the complexities of the algorithm and the problem.

Each agent has its own PTM generator and a pointer attached to it.
The pointer starts with the first bit of the PTM sequence and
after each consultation moves one step further, so that the next
bit of the PTM sequence can be used, if the strategy is
unsuccessful. The control is realized by changing the parameter
$v$ from $0$ to $1 + \gamma$. For the limiting values of the
parameter $v$, the algorithm ${\cal A}$ produces the following
dynamics of the agents.

When $v = 0$, then $D_{j}(v) = D_{j}^{+}$. Therefore, the
successful interval $[D_{j}^{-},D_{j}(v)]$ coincides with the
whole interval $[D_{j}^{-},D_{j}^{+}]$ and the last strategy of
each agent is successful at each step $j = 1,...,n-1$. This means
that each agent always uses the random strategy and the strategy
matrix becomes
$$
S = \left(\begin{array}{ccccc}
+1 & +1 & ... & +1 & +1\\
+1 & +1 & ... & +1 & +1\\
 . &  . & ... & .  &  .\\
+1 & +1 & ... & +1 & +1\\
+1 & +1 & ... & +1 & +1\\
\end{array} \right)
$$

In the opposite limit, when $v = 1 +\gamma$, then
$D_{j}(v)<D_{j}^{-}$. Thus, the unsuccessful interval
$(D_{j}(v),D_{j}^{+}]$ covers the whole interval
$[D_{j}^{-},D_{j}^{+}]$. This means that the last strategy of each
agent is always unsuccessful and, according to the rule, at each
step $j = 1,..., n-1$ an agent asks the PTM generator which
strategy to use next. As a result, the binary sequence of each
agent becomes the initial segment of length $(n-1)$ of the PTM
sequence and the strategy matrix is turned to
$$
S = \left(\begin{array}{ccccc}
+1 & -1 & ... & - 1 & + 1\\
+1 & -1 & ... & - 1 & + 1\\
 . &  . & ... & .  &  .\\
+1 & -1 & ... & - 1 & + 1\\
+1 & -1 & ... & - 1 & + 1\\
\end{array} \right)
$$

The role of the parameter $v$ may be seen in terms of
correlations. When $v = 0$, the agents are independent, because
the behavior of an agent does not depend on the others. However,
as the parameter $v$ gets larger, the successful interval gets
smaller, and the behaviors of the agents become more correlated
through the PTM sequence. To stay independent, an agent has to
show a result approaching the current minimum. Consequently, the
rule urges an agent to follow the PTM sequence more strongly and
as a result, the presence of the PTM sequence in the strategy
matrix becomes more evident. When $v = 1 + \gamma$, the whole
strategy matrix consists of the PTM sequences.

By extensive computational experiments, we investigate how the
performance of the algorithm ${\cal A}$ may depend on the
parameter $v$. We tested the problems $P$ = $\{$eil76, eil101,
st70, rat195, lin105, kroC100, kroB100, kroA100, kroD100, d198,
kroA150, pr107, u159, pr144, pr144, pr152, pr226, pr136, pr76,
ts225$\}$ belonging to the benchmark traveling salesman problems
\cite{Reinelt_1}. Let $D_{i}(p,v)$ be the distance travelled by
agent $i = 1,...,N$ for a problem $p \in P$ and a value $v$ of the
parameter. The performance of the algorithm ${\cal A}$ is
characterized by the average distance travelled by $N$ agents
$D(p,v) = \sum_{i = 1}^{N} D_{i}(p,v)/N$, which is sought to be
minimized.

In the experiments, the algorithm ${\cal A}$ is applied to each
problem $p \in P$ with the number of agents $N = 2000$ and values
$v_{i} = i\triangle v, \ i = 0, 1,...,20, \ v = 1.1$ of the
parameter, where $\triangle v = 0.05$. To eliminate randomness,
the computations are repeated in a series of $10$ tests and the
performance functions $\{D_{k}(p,v), \ k = 1,...,10 \}$ are
averaged into $\bar D(p,v)$. Remarkably, for each problem $p \in
P$, it is found that the performance $\bar D(p,v)$ of the
optimization algorithm ${\cal A}$ behaves as a {\it convex
function} of the parameter $v$ with the only optimum at $v^{*}(p)$
(Fig. 1).

\begin{figure}
\includegraphics[width=.46\textwidth]{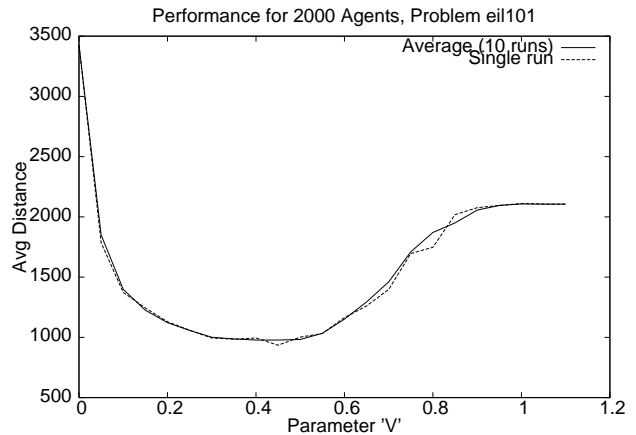}
\caption{\label{fig:zero} The performance of the algorithm ${\cal
A}$ as a convex function of the parameter $v$ for problem ei101.}
\end{figure}

We characterize the performance optimums $\{v^{*}(p), \ p \in P
\}$ by certain quantities of the algorithm and the problem and
interpret the quantities as their complexities. The algorithm
${\cal A}$ is described by the quadratic trace of the
variance-covariance matrix derived from the strategy matrix
\cite{Korotkikh_2}. Namely, for a problem $p \in P$, a set of
strategy matrices $ S_{k}(v^{*}(p)) = \{ s_{ij}(k), i = 1,...,N; \
j = 1,...,n-1 \}, k = 1,...,10 $ is obtained as a result of ten
runs for the value $v^{*}(p)$ of the parameter. For each strategy
matrix $S_{k}(v^{*}(p)), k = 1,...,10$ the variance-covariance
matrix $ V(S_{k}(v^{*}(p))) = \{ V_{ij}(k), \ i, j = 1,...,N \}, i
= 1,...,10, $ and its quadratic trace
$$
tr(V^{2}(S_{k}(v^{*}(p)))) = \sum_{i=1}^{N} \lambda_{ik}^{2}
$$
are computed, where $V_{ij}(k)$ is the linear correlation
coefficient between agents $i = 1,...,N$ and $j = 1,...,N$ and
$\lambda_{ik}, i = 1,...,N, \ k = 1,...,10$ are the eigenvalues of
the variance-covariance matrix $V(S_{k}(v^{*}(p)))$.

The average $tr(V^{2}(S(v^{*}(p))))$ of the traces is used to
describe the complexity
$$
C({\cal A}(p)) = \frac{1}{N^{2}}tr(V^{2}(S(v^{*}(p))))
$$
of the algorithm ${\cal A}$. There is a connection between
$N^{2}/tr(V^{2}(S))$ and the number of KLD (Karhunen-Loeve
decomposition) modes $D_{KLD}$ \cite{Korotkikh_3}. The quantity
$D_{KLD}$ measures the complexity of spatiotemporal data
\cite{Sirovich_1} and a correlation length $\xi_{KLD}$, based on
$D_{KLD}$, can characterize high-dimensional inhomogeneous
spatiotemporal chaos \cite{Zoldi_1}. In our case, it turns more
appropriate to consider the complexity of the algorithm ${\cal A}$
in terms of $tr(V^{2}(S))/N^{2}$, although the complexity is
greater, the smaller $tr(V^{2}(S))/N^{2}$.

We describe the complexity $C(p)$ of the problem $p \in P$ by the
quadratic trace
$$
C(p) = \frac{1}{n^{2}}tr(M^{2}(p)) = \frac{1}{n^{2}}\sum_{i=1}^{n}
\lambda_{i}^{2}
$$
of the distance matrix $ M(p) = \{ d_{ij}/d_{max}, i,j = 1,...,n
\}, $ where $\lambda_{i}, i = 1,...,n$ are the eigenvalues of the
distance matrix, $d_{ij}$ is the distance between cities $i =
1,...,n$ and $j = 1,...,n$ and $d_{max}$ is the maximum of the
distances.

An optimality condition of the algorithm ${\cal A}$ is sought
through a possible relationship between the complexity $C({\cal
A}(p))$ of the algorithm ${\cal A}$ and the complexity $C(p)$ of
the problem $p$. For this purpose, we consider the points with
coordinates $\{x = C(p), y = C({\cal A}(p)), \ p \in P \}$. The
result of the analysis shown in Fig. 2 suggests a possible linear
relationship between the complexities. The regression line is
calculated
$$
y = \alpha x + \beta = 0.67x + 0.33,
$$
\begin{equation}
\label{OC1} \alpha= 0.67 \pm 0.01, \ \ \ \beta = 0.33 \pm 0.01,
\end{equation}
where the standard error of estimate is $0.09$ and the absolute
value of the maximal individual error is $0.05$. The coefficient
of determination of $0.71$ tells that $71$ percent of the
variation in the complexity of the algorithm ${\cal A}$ is
explained by the regression line.

\begin{figure}
\includegraphics[width=.44\textwidth]{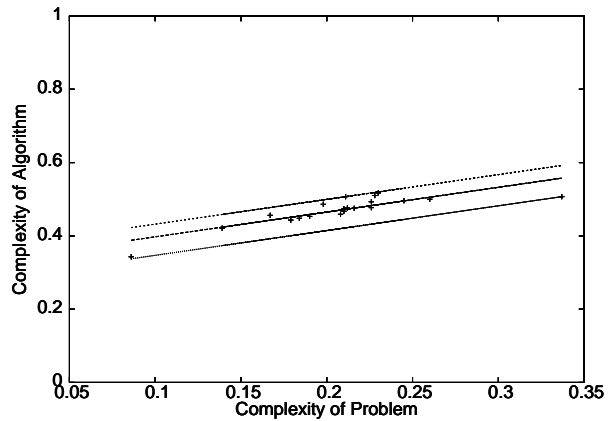}
\caption{\label{fig:zero} The regression line of the relationship
between the complexity of the algorithm ${\cal A}$ and the
complexity of the problem with the bounds including all points.}
\end{figure}

The experiment is repeated a number of times and each time the
results confirm the consistency of the regression line
(\ref{OC1}). Therefore, within the accuracy of the linear
regression, we are able to formulate an optimality condition: {\it
if the algorithm ${\cal A}$ shows its best performance for a
problem $p$, then the complexity $C({\cal A}(p))$ of the algorithm
${\cal A}$ is in the linear relationship with the complexity
$C(p)$ of the problem $p$}
\begin{equation}
\label{OC2} C({\cal A}(p)) = 0.67 \times C(p) + 0.33.
\end{equation}

Computational investigations on the role of the PTM generator for
the algorithm ${\cal A}$ suggest how the optimality condition
(\ref{OC2}) may be extended. For this reason we use a different
algorithm ${\cal B}$, which works exactly in the same manner as
the algorithm ${\cal A}$, except it consults a random generator
instead of the PTM generator. To find a possible connection
between the best performances and the relationship between the
complexities, we compare the algorithms ${\cal A}$ and ${\cal B}$.

First, for each problem $p \in {\cal P}$ the best performance of
the algorithm ${\cal A}$ is compared with the best performance of
the algorithm ${\cal B}$. Fig. 3 shows that the algorithm ${\cal
A}$ demonstrates significantly better results than the algorithm
${\cal B}$ for seventeen problems and results for the other three
are close.

\begin{figure}
\includegraphics[width=.46\textwidth]{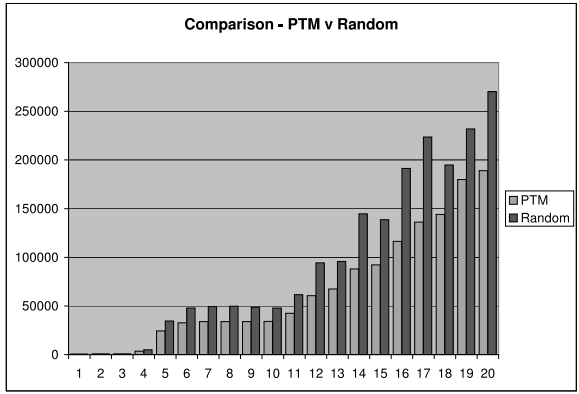}
\caption{\label{fig:zero} The best performances of the algorithm
${\cal A}$ using PTM generator are compared with the best
performances of the algorithm ${\cal B}$ using random generator.
The problems are ordered as in the description of the class $P$.}
\end{figure}

Second, we examine the relationship between the complexities of
the algorithm ${\cal B}$ and the problem (Fig. 4) in the same
manner as we have done for the algorithm ${\cal A}$ (Fig. 2). The
situation deteriorates and the relationship is less consistent for
the algorithm ${\cal B}$ in comparison with the algorithm ${\cal
A}$. Formally, the calculated regression line
$$
y = \alpha'x + \beta' = 0.48x + 0.23,
$$
\begin{equation}
\label{OC3} \alpha' = 0.48 \pm 0.02, \ \ \ \beta' = 0.23 \pm 0.01
\end{equation}
is a less accurate estimator of the relationship between the
complexities of the algorithm ${\cal B}$ and the problem. The
standard error of estimate of $0.18$ and the absolute value of the
maximal individual error of $0.1$ of the regression line
(\ref{OC3}) are doubled in comparison with the regression line
(\ref{OC1}). The coefficient of determination is only $0.26$ in
comparison with $0.71$ for the algorithm ${\cal A}$.

Therefore, computationally, we reveal a connection between the
best performances and the relationship between the complexities.
In particular, the better result of the algorithm ${\cal A}$ (Fig.
3) corresponds to the fact that the relationship between the
complexities of the algorithm ${\cal A}$ and the problem appears
more consistent than it is in the case of the algorithm ${\cal
B}$. In order to suggest potential implications of the optimality
condition (\ref{OC2}) to a wider context, we would like to use
this proposition.

The observed connection allows us to assume a possible sequence of
algorithms whose best performances converge to the optimal
solutions of the problems as the relationships between the
complexities converge to a certain function $g$. Notice, that the
algorithms ${\cal A}$ and ${\cal B}$ may be well among the
elements of the sequence with the algorithm ${\cal B}$ followed by
the algorithm ${\cal A}$. According to the experiments, it seems
likely that the function $g$ may be a linear function.

\begin{figure}
\includegraphics[width=.44\textwidth]{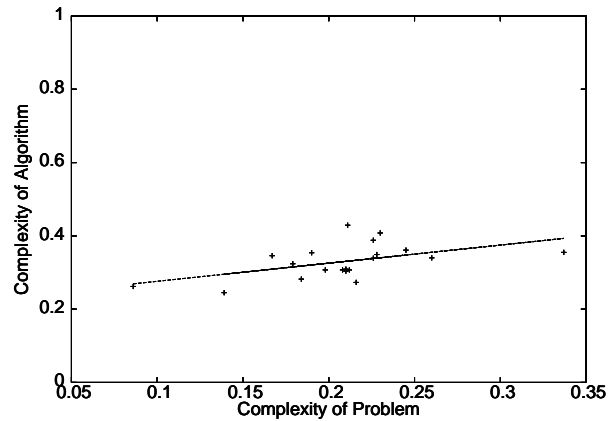}
\caption{\label{fig:zero} The regression line of the relationship
between the complexity of the algorithm ${\cal B}$ and the
complexity of the problem.}
\end{figure}

The possible convergence of the sequence would promise an
algorithm ${\cal G}$ to link the optimal performance and the
relationship between the complexities. As a result, the best
performance of the algorithm ${\cal G}$ for any problem could
actually provide the optimal solution. An optimality condition of
the algorithm ${\cal G}$ may extend the optimality condition of
the algorithm ${\cal A}$ and explain the optimal performance in
terms of the relationship between the complexities: if the
algorithm ${\cal G}$ shows its best performance for a problem $p$,
and thus finds the optimum solution, then the complexity $C({\cal
G}(p))$ of the algorithm is in the relationship $C({\cal G}(p))=
g(C(p))$ with the complexity $C(p)$ of the problem.

The optimality condition of the algorithm ${\cal G}$ may suggest a
new approach to optimization that would be primarily concerned
with the efficient control of the algorithm's complexity in
matching the complexity of the problem. For example, if the
algorithm ${\cal G}$ could potentially find the optimal solution
to a problem $p$ of complexity $C(p)$ under the condition $C({\cal
G}(p)) = g(C(p))$, would it then be possible to obtain the optimal
solution by a different algorithm working with the same complexity
for the problem?

The ability to evaluate the complexity of the problem before the
actual computation would be beneficial. In this case, the finding
of the optimal solution could be connected with the tuning of the
algorithm's complexity in order to match the already known
complexity of the problem. Experimentally, it is observed that for
any problem the performance of the algorithm ${\cal A}$ is a
convex function of the parameter. As a result, the best
performance of the algorithm ${\cal A}$ for the problem can be
efficiently obtained by the minimization of this convex function.
Therefore, it would be important to understand whether the
performance of the algorithm ${\cal G}$, as the function of its
complexity, may behave in a similar way to provide us with
efficient means for finding optimal solutions.

In conclusion, we have presented an optimization algorithm as a
complex system. For any problem tested, the performance of the
algorithm has been controlled as a convex function with a single
optimum. By the characterization of the performance optimums, an
optimality condition of the algorithm has been proposed. The
optimality condition has revealed that the relationship between
certain quantities of the complex system and the problem may
determine the optimal performance. The result provides a
computational evidence to the possibility of an optimality
condition of complex systems.

This work was supported by CQU Research Advancement Awards Scheme
grant no. IN9022 and Faculty of Informatics and Communication
Research Grant Scheme no. FRG00406.


\begin{thebibliography}{00}

\bibitem{Holland_1}
J. H. Holland {\it Emergence: From Chaos to Order}
(Addison-Wesley, Reading, Massachusetts, 1998); Y. Bar-Yam {\it
Dynamics of Complex Systems} (Westview Press, 1997); S. Kauffman
{\it At Home in the Universe} (Oxford University Press, New York,
1995).
\bibitem{Ball_1} P. Ball, Nature {\bf 402}, c73 (1999),
and references therein.
\bibitem{Reinelt_1} G. Reinelt {\it TSPLIB Version 1.2
$\lbrack$ online $\rbrack$}
(ftp://ftp.wiwi.unifrankfurt.de/pub/TSPLIB 1.2) $\lbrack$ Accessed
28/11/2000$\rbrack$.
\bibitem{Korotkikh_1} V. Korotkikh {\it A Mathematical Structure
for Emergent Computation} (Kluwer, Dordrecht, 1999).
\bibitem{Feigenbaum_1} M. Feigenbaum, Los Alamos Sci. {\bf 1},
4 (1980).
\bibitem{Allouche_1} J. Allouche and M. Cosnard, in {\it Dynamical Systems
and Cellular Automata} (Academic Press, 1985).
\bibitem{Korotkikh_2} G. Korotkikh and V. Korotkikh, in
{\it Optimization and Industry: New Frontiers}, edited by P.
Pardalos and V. Korotkikh (Kluwer, Dordrecht, 2003).
\bibitem{Korotkikh_3} V. Korotkikh and G. Korotkikh, in
{\it Quantitative Neuroscience}, edited by P. Pardalos, C.
Sackellares, P. Carney and L. Iasemidis (Kluwer, Dordrecht, 2004).
\bibitem{Sirovich_1} L. Sirovich and A.E. Deane, J. Fluid Mech.
{\bf 222}, 251 (1991); S. Ciliberto and B. Nikolaenko, Europhys.
Lett. {\bf 14}, 303 (1991); R. Vautard and M. Ghil, Physica
(Amsterdam) {\bf 35D}, 395 (1989).
\bibitem{Zoldi_1} S.M. Zoldi and H.S. Greenside, Phys. Rev. Lett.
{\bf 78}, 1687 (1997).

\end{thebibliography}
\end{document}